\documentclass[sigconf,nonacm]{acmart}
\input{preface.tex}

\AtBeginDocument{%
  \providecommand\BibTeX{{%
    \normalfont B\kern-0.5em{\scshape i\kern-0.25em b}\kern-0.8em\TeX}}}

\setcopyright{acmcopyright}
\copyrightyear{2020}
\acmYear{2020}
\acmDOI{10.1145/1122445.1122456}

\acmConference[CIKM'21]{SIGKDD}{November 1--5, 2021}{}
\acmPrice{15.00}
\acmISBN{978-1-4503-XXXX-X/18/06}

\begin{document}

\title{
Accelerating COVID-19 research with graph mining and transformer-based learning}


\author{Ilya Tyagin}
\orcid{0000-0002-0249-1072}
\affiliation{%
  \institution{Center for Bioinformatics and Computational Biology \\University of Delaware}
  \city{Newark}
  \state{DE}
  \postcode{19713}
  \country{USA}
  }
\email{tyagin@udel.edu}

\author{Ankit Kulshrestha}
\orcid{}
\affiliation{%
  \institution{Computer and Information Sciences\\University of Delaware}
  \city{Newark}
  \state{DE}
  \postcode{19716}
  \country{USA}
  }
\email{akulshr@udel.edu}

\author{Justin Sybrandt}
\authornote{Now with Google Brain. Contact: jsybrandt@google.com.}
\orcid{0000-0001-5073-0122}
\affiliation{%
  \institution{School of Computing\\Clemson University}
  \streetaddress{821 McMillan Rd.}
  \city{Clemson}
  \state{SC}
  \postcode{29630}
  \country{USA}
  }
\email{jsybran@clemson.edu}

\author{Krish Matta}
\affiliation{%
  \institution{Charter School of Wilmington}
  \city{Wilmington}
  \state{DE}
  \postcode{19807}
  \country{USA}
  }
\email{matta.krish@charterschool.org}

\author{Michael Shtutman}
\affiliation{%
  \institution{Drug Discovery and  Biomedical Sciences\\ University of S. Carolina}
  \streetaddress{715 Sumter Street}
  \city{Columbia}
  \state{SC}
  \postcode{29208}
  \country{USA}
  }
\email{shtutmanm@sccp.sc.edu}

\author{Ilya Safro}
\affiliation{%
  \institution{Computer and Information Sciences\\University of Delaware}
  \city{Newark}
  \state{DE}
  \postcode{19716}
  \country{USA}
  }
\email{isafro@udel.edu}

\begin{abstract}
In 2020, the White House released the, ``Call to Action to the Tech Community on New Machine Readable COVID-19 Dataset,'' wherein artificial intelligence experts are asked to collect data and develop text mining techniques that can help the science community answer high-priority scientific questions related to COVID-19. The Allen Institute for AI and collaborators announced the availability of a rapidly growing open dataset of publications, the COVID-19 Open Research Dataset (CORD-19). As the pace of research accelerates, biomedical scientists struggle to stay current. To expedite their investigations, scientists leverage hypothesis generation systems, which can automatically inspect published papers to discover novel implicit connections. We present automated general purpose hypothesis generation systems \sysnamec and \sysnamegp for COVID-19 research. The systems are based on the graph mining and transformer models. The systems are massively validated using retrospective information rediscovery and proactive analysis involving human-in-the-loop expert analysis. Both systems achieve high-quality predictions across domains in fast computational time and are released to the broad scientific community to accelerate biomedical research. In addition, by performing the domain expert curated study, we show that the systems are able to discover on-going research findings such as the relationship between COVID-19 and oxytocin hormone.\\ 
\noindent {\bf Reproducibility:} All code, details, and trained models are available at \url{https://github.com/IlyaTyagin/AGATHA-C-GP}.
\end{abstract}

%
%

%
%


\maketitle

\section{Introduction}
\label{sec:introduction}

Development of vaccines for COVID-19 is a major triumph of modern medicine and humankind's ability to accelerate scientific research.
While we are all hoping to see large-scale positive changes from fast mass adoption of the existing vaccines, there remain significant open research questions around COVID-19.
The scientific community has a responsibility to do everything possible to block the ongoing transmission of the dangerous virus and \emph{accelerate research to mitigate its consequences}.
We present the following automated knowledge discovery system in order to propose new tools that could compliment the existing arsenal of techniques to accelerate 
biomedical and drug discovery research for events like COVID-19.

The COVID-19 pandemic became one of the most important events in the information space since the end of 2019. The pace of published scientific information is unprecedented and spans all resolutions, from the news and 
pop-science articles 
to drug design at the molecular level. The pace of scientific research has already been a significant problem in science for years \cite{spangler2015accelerating}, and under current circumstances this factor becomes even more pronounced. Several thousands papers are being added \emph{weekly} to CORD-19\footnote{https://www.semanticscholar.org/cord19} (the dataset of publications related to COVID-19) and even more in MEDLINE\footnote{https://www.nlm.nih.gov/bsd/stats/cit\_added.html}. As a result, groups working on similar problems may not be immediately aware of the other's findings, which can lead to inefficient investments and production delays.

\ifthesis

    \begin{figure}
        \centering
        \includegraphics[width=0.9\linewidth]{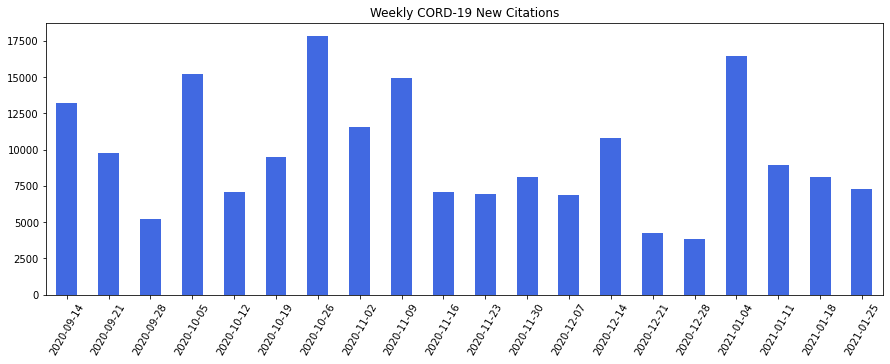}
        \caption{
          \label{fig:cord_papers_rate}
          Number of new citations per week in CORD-19 dataset. 
        }
    \end{figure}

    Under normal circumstances, the MEDLINE database of biomedical citations receives approximately 950,000 new papers per year. Currently this database indexes 31 million total citations.
    This pace challenges traditional research methods, which often rely on human intuition when searching for relevant information. As a result, 
    the demand for modern AI solutions to help with the automated analysis of scientific information  is incredibly high. 
    For instance, the field of drug discovery has explored a range of AI analytical tools to expedite new treatments~\cite{gopalakrishnan2019survey}. 
    Designing lab experiments and finding candidate chemical compounds is a costly and long-lasting procedure, often taking years. To accelerate scientific discovery, researchers came up with a family of strategies to utilize public knowledge from databases like MEDLINE that are available through the National Institute of Health (NIH), which facilitate \emph{automated hypothesis generation} (HG) also known as literature-based discovery. Undiscovered public knowledge, information that is \textit{implicitly} present within available literature, but is not yet \textit{explicitly} known by an individual who can act on that information, represents the target of our work.
    
\fi

Although, there are quite a few hypothesis generation (HG) systems \cite{gopalakrishnan2019survey} including those we have previously proposed~\cite{sybrandt2017,sybrandt2020agatha}, \emph{none of them is currently COVID-19 customized and available in the {\bf open domain} to massively process related queries}. In addition to the traditional  requirements for HG systems, such as high-quality results of hypotheses, interpretability and availability for broad scientific community, a specific demand for COVID-19 data analysis requires: (1) customization of the vocabulary and other logical units such as subject-verb-object predicates; (2) customization of the training data that in the reality of urgent research contains a lot of controversial and incorrect information; (3) multiple models for different information resolutions (e.g., microscopic for drug design, and macroscopic for the population related conclusions); and (4) validation on the on-going domain-specific discovery. 

\noindent{\bf Our contribution:} In this work we bridge this gap by releasing, \sysnamec and \sysnamegp, reliable and easy to use HG systems that demonstrate state-of-the art performance and validate their inference capabilities on both COVID-19 related and general biomedical data. To make them closely related to different goals of COVID-19 research, they correspond to micro- (AGATHA-C, for COVID-19) and macroscopic (AGATHA-GP, for general purpose) scales of knowledge discovery. Both systems are trained on all existing biomedical literature available through NIH and CORD-19 and able to process any queries to connect biomedical concepts but \sysnamec exhibits better results on the molecular scale queries, e.g., those that are relevant to drug design, and \sysnamegp works better for general queries, e.g., establishing connections between certain profession and COVID-19 transmission. As it will be explained later, we  emphasize that AGATHA is not a traditional \emph{information retrieval} system that effectively searches for \emph{existing} information and thus cannot be compared to them. Instead, AGATHA generates novel hypotheses.

Both systems are the next generation of the AGATHA knowledge network mining transformer model \cite{sybrandt2020agatha}. \emph{They substantially improve the quality of the previous AGATHA by introducing new information layers into multi-layered semantic knowledge network pipeline, and expanding new information retrieval techniques that facilitate inference.} 
We present the deep learning transformer-based \pymoliere-C/GP models trained with up-to date datasets and provide easy to use interface to broad scientific community to conduct COVID-19 research. We validate the system via candidate ranking \cite{sybrandt2018a,sybrandt2020agatha} using very recent scientific publications containing findings absent in the training set.
While the original AGATHA has demonstrated state-of-the-art performance for the time of its release, AGATHA and other systems were found to perform with notably lower quality on extremely rapidly changing COVID-19 research. We demonstrate a remarkable improvement on different types of queries with very fast query process that allows massive validation. In addition, we demonstrate that the proposed system can identify recently uncovered gene (BST2) and hormone (oxytocin and melatonin) relationships to COVID-19, using only papers published before these connections were discovered.
\ifthesis
\else
    More technical details are available in our extended preprint version of the paper at arXiv~\cite{tyagin2021accelerating}.
\fi

\ifthesis
    \noindent {\bf Reproducibility:} All code, details, and pre-trained models are available at \url{https://github.com/IlyaTyagin/AGATHA-C-GP}.
\fi

\section{Background and Related Work}
\label{sec:background}

\ifthesis

    The HG field has been present in information sciences for several decades. The first notable approach was proposed by Swanson et al. in 1986~\cite{swanson1986fish}, which is called the A-B-C model. The concept of A-B-C model is to discover intermediate (B) terms which occur in titles of publications for both terms A (source) and C (target). In their experiments, Swanson et al. discovered an implicit connection between Raynauld's syndrome (term A) and fish oil (term C) through blood viscosity (term B), which was mentioned in both sets. 
    The hypothesis that fish oil can be used  for patients with Raynaud’s disease was experimentally confirmed several years later~\cite{pmid2536517}. The key idea of the proposed method is that all fragmented bits of information are explicitly known, but their implicit relationships is what HG systems are aimed to uncover. 

    We note the difference between HG (not only the A-B-C models) and traditional information retrieval. The information retrieval techniques which represent the vast majority of biomedical literature based discovery systems (including those for COVID-19) are trained and (what is even more important) validated to retrieve \emph{existing  information} whereas the HG techniques predict \emph{undiscovered knowledge} and thus must be massively validated on it. The HG validation requires training the system strictly on historical data rather than sampling it over the entire time. 

    The advances in machine and deep learning transformed the algorithms of HG systems that are now able to process much larger information volumes demonstrating higher quality predictions. 

\fi


A number of works have been proposed to organize the CORD-19 literature into structured graphs for different purposes 
\ifthesis
    . For instance, Basu \emph{et al.}~\cite{basu} propose ERLKG - a knowledge graph built on CORD-19 with entities corresponding to gene/chemical/disease names and the edges forming relations between the concept. They use a fine tuned SciBERT model for both entity and relation extraction. The main purpose of the knowledge graph is to predict a link between a given chemical-disease and chemical-protein pair using a trained GCN autoencoder ~\cite{kipf2017semi} approach. In another similar work, Oniani \emph{et al.}~\cite{oniani} build a co-occurrence network on a subset of CORD-19 with the edges corresponding to either gene-disease, gene-mutation or chemical-disease type. The network is then embedded into latent space using a node2vec walk. Link prediction is performed on the nodes by training different classical machine learning algorithms. 
    \else
    \cite{basu,oniani}.
\fi
A major shortcoming of these approaches is that they are limited to either specific kind of entities or relations or both and as a result not only the scope of possible new literature is narrowed but a lot of additional useful knowledge is filtered out of the system. 
\ifthesis
    In contrast, our system does not limit itself to specific entity or relation type and is able to capture much more information from the same corpus.
\fi

A major interest of constructing knowledge graphs is to allow medical researchers to re-purpose existing drugs for treating COVID-19. Zhang \emph{et al.}~\cite{zhang2020drug} develop a system that uses combined semantic predicates from SemMedDB and CORD-19 (extracted using SemRep) to recommend drugs for COVID-19 treatment. To improve the predications from CORD-19, the authors fine tune various transformer based models on a manually annotated internal dataset. 
\ifthesis
    Their resulting knowledge graph consists of 131,555 nodes and 2,558,935 edges. Our work on the other hand utilizes similar technologies and produces a bigger graph with 287,356,836 nodes and 13,500,291,256 edges. Moreover, we do not post-process extracted relations from SemRep and are still able to achieve a higher RoC metric.
\fi
 
 The most similar to our current work is the system proposed by~\cite{Nordon2019SeparatingWF}. The authors use Electronic Medical Record (EMR) to generate candidates for drug repurposing and then use a knowledge graph constructed from MEDLINE documents to validate those candidates. In context of recent diseases like COVID-19 for which EMRs are not as readily available, the system will be limited in their candidate generation. Moreover, validation using only SemMedDB publication data may yield subpar results since it is not updated as frequently. In contrast, our method incorporates new sources of literature during the graph construction phase itself and can be readily adapted to new and emerging challenges in medicine.  Other systems that are built for COVID-19 drug discovery include systems by~\cite{martinc} and Kinderminer~\cite{pmid28815126}. The former tool uses fine-tuned SciBERT model to generate contexualized embeddings given an initial seed set of words and the latter system uses a keyword co-count algorithm to propose candidates for COVID-19. We observe that our graph contains a larger variety of data sources data than any of these tools and thus can produce broader set of hypotheses.

\ifthesis

    The vastness of COVID-19 literature also spurned the need for having systems that could allow researchers and base users alike to get their COVID-19 queries answered. Systems like CKG (Wise \emph{et al.})~\cite{wise2020covid19} and SciSight (Hope \emph{et al.})~\cite{hope2020scisight} currently provide this functionality. While we do aim to provide an easy to use web-framework for medical researchers, the scope of the aforementioned systems is beyond the scope of our work. Unfortunately, no existing system out of those that are trained to accept terms related to COVID-19 or SARS-CoV-2 provided an open access for massive validation for a fair comparison with or was able to be tested in multiple domains like \sysnamec.

\fi

The lack of broader applicability of systems like these in the situation with COVID-19 pandemic demonstrates that several major issues exist and require immediate attention:\\
\noindent (1) Most of the existing HG systems are domain-specific (e.g., gene-disease interactions) that is usually expressed in limiting the processed information (e.g., significant filtering vocabulary and papers to a specific domain in probabilistic topic modeling  \cite{wang2011finding});\\
\noindent (2) A proper validation of HG system remains a technical problem because multiple large-scale models have to trained with all heterogeneous data carefully eliminated several years back;\\
\noindent (3) Moreover, a large number of HG systems are not massively validated at all except of very old findings rediscovery \cite{smalheiser2017rediscovering} or demonstrating of just a few proactive examples in humanly curated investigation; and \\
\noindent (4) Interpretability and explainability of generated hypotheses remains a major issue.

\noindent{\bf The UMLS Metathesaurus}~\cite{Bodenreider04theunified} is the NIH database containing information about millions of concepts (both medical and general) and their synonyms. 

\ifthesis
    Metathesaurus accumulates information about its entries from more than 200 different vocabularies allowing to map and connect concepts from different terminologies. Metathesaurus also keeps metadata about the concepts such as semantic types and their hierarchy. 
    The core unit of information in UMLS is the concept unique identifier, or CUI. CUI is a codified representation of a specific term, which includes its different atoms (spelling variants or translations of the term on other languages), vocabulary entries, definitions and other metadata.
\fi


\noindent{\bf SemRep}~\cite{arnold2015semrep} is a software kit developed by NIH for extraction of semantic predicates (subject-verb-object triples) from the provided corpus. It also allows to extract entities not involved in any semantic predicate, if the corresponding option is selected. 

\ifthesis
    The official example of  possible SemRep output is: INPUT = ``We used hemofiltration to treat a patient with digoxin overdose that was complicated by refractory hyperkalemia.'',  OUTPUT = ``Hemofiltration - TREATS - Patients; Digoxin overdose - PROCESS\_OF - Patients;  hyperkalemia-COMPLICATES-Digoxin overdose;  Hemofiltration - TREATS (INFER) - Digoxin overdose''.
    SemRep handles word sense disambiguation and performs terms mapping to the corresponding CUIs from UMLS metathesaurus.
\fi

\noindent{\bf ScispaCy}~\cite{neumann2019scispacy} ScispaCy is a special version of spaCy maintained by AllenAI, containing spaCy models for processing scientific and bio-related texts. ScispaCy models are trained on different sources, such as PMC-pretrained word2vec representations, MedMentions Entity linking Dataset and so on. 
SciSpacy can handle various NLP tasks, such as NER, dependency parsing and POS-tagging, where achieves state of the art performance.

\noindent{\bf SciBERT}~\cite{beltagy2019scibert} is a BERT-like transformer pretrained language model, where full-text scientific papers were used as a training dataset. Embeddings are learned in a word-piece fashion, which makes them capture the relationships between not only words in a sentence, but also between word parts in each word.

\noindent{\bf FAISS}~\cite{johnson2017faiss} is a library for fast approximate clustering and similarity search between dense vectors. 
It scales to the huge datasets that do not fit in RAM and can be used in a distributed fashion. FAISS is used in our pipeline to perform $k$-means clustering of PQ-quantizated sentence vectors to generate $k$-nearest neighbor edges for similar sentences (nodes) in knowledge network.

\noindent{\bf PTBG}~\cite{lerer2019biggraph} (stands for PyTorch BigGraph) is a high-performance graph embedding system allowing distributed training. 
It was designed to handle large heterogeneous networks containing hundreds of millions of nodes of different types and billions of typed edges. Distributed training is achieved by computing embeddings on disjoint node sets.

\noindent{\bf AllenNLP Open Information Extraction}. AllenNLP~\cite{Gardner2017AllenNLP} is a powerful library developed by AllenAI that uses  PyTorch backend to provide deep-learning models for various natural processing tasks. 
Specifically, AllenNLP Open Information Extraction provides a trained deep bi-LSTM model for extracting predicates from unstructured text. An API is provided for running inference in both single sentence and batch modes.

\noindent{\bf CORD-19 dataset}~\cite{Wang2020CORD19TC} was released as a response to the world's COVID-19 pandemic to help data science experts and researchers to tackle the challenge of answering the high priority scientific questions. It updates daily and was created by the Allen Institute for AI in collaboration with Microsoft Research, NLM, IBM and other organizations. At the time of writing this paper it contains over 400.000 scientific abstracts and over 150.000 full-text papers about coronaviruses, primarily COVID-19.

\noindent{\bf MEDLINE} is a database of NIH that includes almost 31 million citations (as of 2021) of scientific papers related to the biomedical and related fields. Some of the citations are provided with MeSH (Medical Subject Headings) terms and other metadata. MEDLINE is one of the largest and well-known resources for biomedical text mining.

\section{Pipeline Summary}
\label{sec:methods}

We briefly summarize the \pymoliere{} semantic graph construction pipeline. It is described in greater details  in~\cite{sybrandt2020agatha}.

\noindent{\bf Text pre-processing}. 
The input for our system is \textit{a corpora of scientific citations} from the MEDLINE and CORD-19 datasets. These files contain titles and abstracts for millions of biomedical papers. We filter non-English documents, using the FastText Language Identification model~\cite{joulin2016fasttext} if the language is not provided.
After that we split all abstracts into sentences and process all sentences with ScispaCy library. From each sentence we extract POS-annotated lemmas, entities and perform $n$-gram mining, where $n \in [2,3,4]$ and $n$-grams are composed of frequently co-occurring lemmas. Additionally, we associate all sentences with any relevant metadata, such as the MeSH/UMLS~\cite{pmid14681409} keywords provided along with the citation.

\begin{figure}
    \centering
    \includegraphics[width=0.9\linewidth]{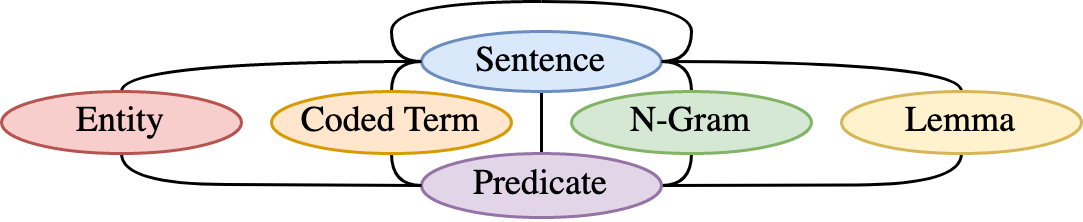}
    \caption{\pymoliere{} multi-layered graph schema.}
    \label{fig:graph_schema}
\end{figure}

\noindent{\bf Semantic Graph Construction}. 
We construct a semantic graph containing different types of nodes, namely, sentences, entities, coded terms (from UMLS and MeSH), $n$-grams, lemmas, and predicates following the schema depicted in Figure~\ref{fig:graph_schema}. Edges between sentences are induced from the nearest-neighbors network of sentence embeddings. We also include an edge between two sentences that appear sequentially within the same abstract, counting the title as the first sentence. Other edges can be inferred directly from the recorded metadata. For instance, the node representing the entity ``COVID-19'' is connected to every sentence and predicate that discuss COVID-19.

\noindent{\bf NLM UMLS implementation}. The prior \pymoliere{} semantic network only includes UMLS terms that appear in SemMedDB predicates~\cite{KilicogluSFRR12} which is a major limitation. In this work we enrich the ``Coded Term'' layer by introducing an additional preprocessing phase wherein we run the SemRep tool with full-fielded output option ourselves \textit{on the entire input corpora}. This phase would be necessary as CORD-19 and most recent MEDLINE citations are not represented within slowly updated SemMedDB. However, we find that we can substantially increase the quality of recovered terms by applying these tools ourselves. By doing that we not only enrich the ``Coded Terms" semantic network layer, but also introduce a significant number of uncovered previously semantic predicates. 

\ifthesis
    It happens because SemMedDB is a cumulative database, having various citations in the database processed over many years with various versions of SemRep and various UMLS releases available at different time periods.
    
    To illustrate what was just said, let us consider the following example (PMID: 20109154): \textit{"The results showed that V. cholerae O395 and also other related enteric pathogens have the essential CASS components (CRISPR and cas genes) to mediate a RNAi-like pathway."} The current SemRep version extracts the following predicate: \texttt{CRISPR-AFFECTS-RNAi}, while SemMedDB does not contain any predicates for this sentence. The year of publication of the corresponding paper is 2009, but CRISPR term (C3658200) did not exist in the UMLS metathesaurus on or before 2012, that is why at the time of adding this citation to SemmedDB CRISPR-involved relation could not be identified.
\fi

\noindent{\bf Graph Embedding.} The resulting semantic graph is a large undirected heterogeneous network, where each node has its own type (as shown in Figure \ref{fig:graph_schema}) and each edge between nodes with types $u$ and $v$ corresponds to type $uv$. At this point we additionally clarify that the constructed network is \textit{not a traditional homogeneous knowledge graph.} 
We embed the network using a heterogeneous technique that captures node similarity through a \textit{biased transformed dot product}. By explicitly including a bias term for each node, we capture a concepts overall affinity within the network that is critical for such general terms as ``coronavirus.'' By learning transformations between each pair of node types (e.g., between sentences and lemmas), we enable each type to occupy embedding spaces with differing characteristics. Specifically, we fit an embedding model that optimizes the following similarity measure:
\begin{equation}
\mathcal{S}(u, v) = \hat{u}_1 + \hat{v}_1 + T^{uv}_1 + \sum_{i=2}^d \hat{u}_i(\hat{v}_i + T^{uv}_i),
\end{equation}
\noindent where $d$ - space dimensionality, $u, v$ are nodes in the semantic graph with embeddings $\hat{u}, \hat{v}$, and $T^{uv}$ is the directional transformation vector between nodes of $u$'s type to nodes of $v$'s.

\ifthesis
    We use the PyTorch BigGraph heterogeneous graph embedding library to learn $d = 512$ dimensional vector embeddings for
    each node of our large semantic graph. While fitting embeddings ($\hat{u}$) and transformation vectors ($T^{uv}$), we represent each edge of the semantic graph as two directed edges. These learned values are optimized using softmax loss, where the similarity for one edge is compared against the similarities of 100 negative samples.
\fi

\noindent{\bf Ranking Semantic Predicates (Transformer model).} After we obtain embeddings per node in the semantic graph, we train \pymoliere{} system ranking model. This model is trained to rank published subject-object pairs above randomly composed pairs of UMLS concepts (negative samples). Two coded terms, along with a fixed-size random subsample of predicates containing each term are input to this model. Graph embeddings for each term and predicate are fed into stacked transformer encoder layers, which apply multi-headed self-attention across the embedding set. The last set of encodings are averaged and the result is projected to the unit interval, forming a scalar prediction for the input's ``plausibility.''

\ifthesis
    Formally, the model to evaluate term pairs is defined as:
    \begin{equation}
    \begin{aligned}
    f(x, y) &= g\left(\begin{bmatrix}
    \hat{x} & \hat{y}
    & \hat{x'_1} \ldots \hat{x'_k}
    & \hat{y'_1} \ldots \hat{y'_k}
    \end{bmatrix}\right)\\
    g(X) &= \text{sigmoid}(\mathcal{M}\Theta) \\
    \mathcal{M} &= \frac{1}{|X|}\text{ColSum}\left(\mathcal{E}_N(\text{FeedForward}(X))\right) \\
    \mathcal{E}_0(X) &= X \\
    \mathcal{E}_{i+1}(X) &= \text{LayerNorm}\left(\text{FeedForward}(\mathcal{A}(X)) + \mathcal{A}(X)\right)\\
    \mathcal{A}(X) &= \text{LayerNorm}\left(\text{MultiHeadAttention}(X) + X\right), \\
    \end{aligned}
    \end{equation}
    where each $x'$ and $y'$ are randomly sampled from the neighborhoods of $x$ and $y$ respectively, and each $\hat{\cdot}$ denotes the graph embedding of the given node. Furthermore, $\Theta$ represents a free parameter, which is fit along with parameters internal to each FeedForward and MultiHeadAttention layer, following the standard conventions for each.
    
    The above model is fit using margin ranking loss, where predicates from the training set are compared against a large set of negative samples. Additional details pertaining to specific optimization choices surrounding this model are present in the work originally proposing this model~\cite{sybrandt2020agatha}.
\fi

\section{Augmenting Semantic Predicates with Deep Learning}
\label{sec:dl_predicate_extraction}

\begin{figure}
    \centering
    \includegraphics[width=\linewidth]{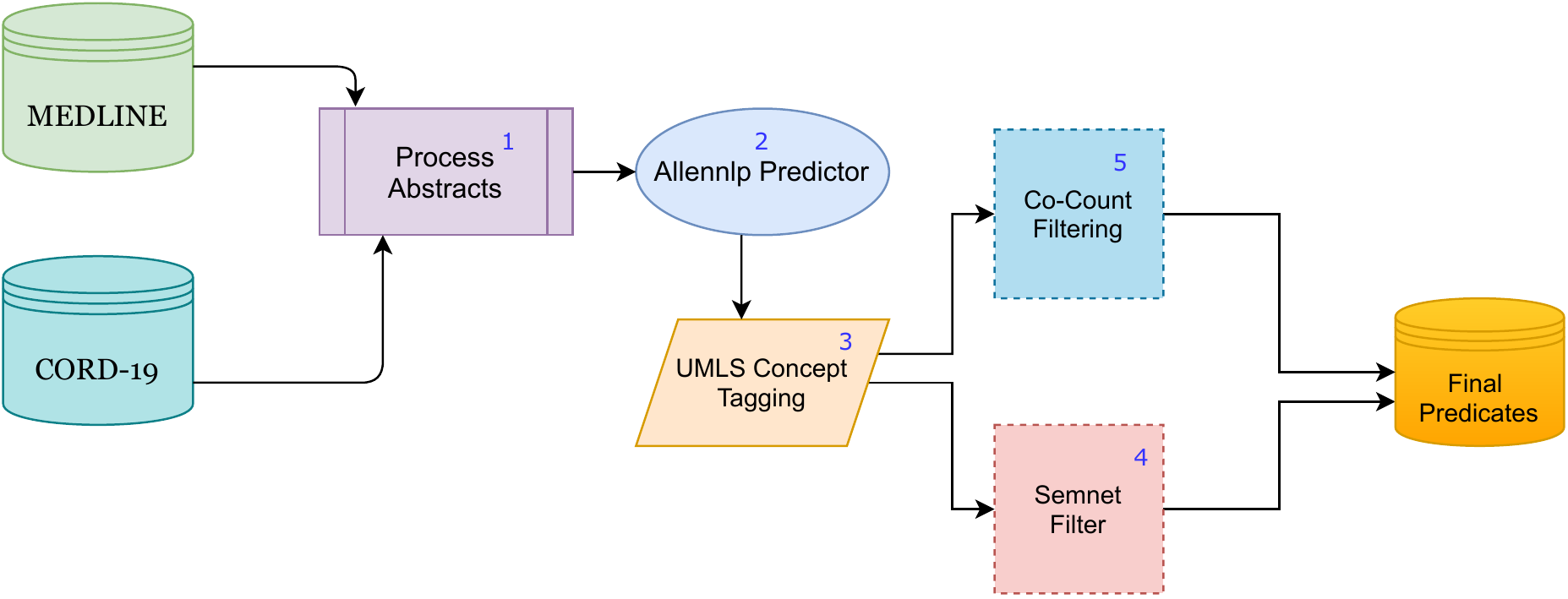}
    \caption{
        Predicate Extraction pipeline with Deep Learning based Open IE system. 
    }
    \label{fig:dl_predicate_extraction}
\end{figure}

We used \texttt{SemRep} predicate extraction system in the first system, \sysnamec, to extract predicates from the abstracts. However, \texttt{SemRep} relies on expert coded rules and heuristics to extract biomedical relations leading to significantly fewer predicates for training. Thus, in order to augment the predicates (for the second system, \sysnamegp\nolinebreak) we decided to use a deep learning based information extraction system by Stanvosky \emph{et al.}~\cite{Stanovsky2018NAACL}. Figure~\ref{fig:dl_predicate_extraction} shows our overall predicate extraction pipeline. We index our pipeline by box numbers and describe them below.
\ifthesis
\noindent \textbf{Abstract Pre-processing [Box 1]} . The input for the proposed semantic predicate extraction system is the output files generated by \texttt{SemRep} tool with full-fielded output option enabled, obtained from the preprocessing stage described in Sec.~\ref{sec:methods}. As it was mentioned previously, \texttt{SemRep} system extracts not only semantic triples, but also maps entities found in the input corpus to their corresponding UMLS concept IDs, this is the data which is used for the following method. The initial set of records includes the sentence raw texts and extracted from them UMLS terms and  is augmented throughout the pipeline making it easier to extract final predicates for downstream training.
\else
\noindent \textbf{Abstract Pre-processing Box [Box 1]}. We use \texttt{Semrep} tool described in previous sections to process the abstracts and mine information about the biomedical entities and potential semantic relations. This information is processed into a record-like data structure and is augmented throughout the rest of the pipeline.
\fi

\ifthesis
\noindent \textbf{Raw Predicate Extraction [Box 2 \& 3]}. We use a pre-trained instance of \texttt{RnnOIE} \cite{Stanovsky2018NAACL} provided as an API by AllenNLP. The model was trained on the \texttt{OIE2016} corpus. At a high level the model aims to learn a joint embedding of individual words and their corresponding Beginning-Input-Output (BIO) tags. The output of the model is a probability distribution over the BIO tags. During inference the model selects specific phrases and groups them into \texttt{ARG0}, \texttt{V}, \texttt{ARG1} tags. By convention, we treat \texttt{ARG0} as the subject and \texttt{ARG1} as the object in a subject-verb-object tuple. To speed up processing and scale it to thousands of abstracts, we leverage model-parallelism across different machines and run batch-mode inference on chunks of abstracts. Once the model predictions have been extracted  we extract the phrases with relevant tags into raw predicates and augment them in the record. A subsequent filtering is performed by extracting the terms matching with previously detected UMLS concepts in the sentence.
\else 
\noindent \textbf{Raw Predicate Extraction [Box 2 \& 3]}. We use the model described in~\cite{Stanovsky2018NAACL} as the deep learning model to extract semantic predicates. The model is provided as a prediction endpoint by Allennlp and is trained to predict the \emph{beginning-input-output}(BIO) tags for a particular sentence and classify them into subject, verb and object. We extract and match the UMLS concepts contained in the phrases and store these as ``raw predicates" for further filtering.
\fi 


\ifthesis
\noindent \textbf{Semnet Filtering [Box 4]} Using a general purpose \texttt{RnnOIE} model has it's own challenges. During processing we noted that a lot of raw predicates were either too general or contained too little meaning to be useful for training a prediction model. To overcome this challenge we designed a corrective filter to reduce noise and retain most useful predicates. We call this filter the \emph{semnet filter}.

Each UMLS concept has an associated semantic type (e.g., COVID-19 has an associated semantic type of \texttt{dsyn} (disease)). This is useful for summarizing large set of diverse text concepts into smaller number of categories. We used the metadata from semantic types to construct two networks - a semantic network and a hierarchical network. The semantic network consists of semantic types as nodes and the edges imply a corresponding direct relation between them.  The hierarchical network is a network of a semantic type connected to its more general semantic types. For example, a semantic type \texttt{dsyn} (disease) is more generally associated with a \texttt{biof} (biological function) or a \texttt{pathf} (pathological function). In order to filter a predicate, all edges emanating from the subject's semantic types are computed on a per-predicate basis. These edges also include any specific-general concept relationships. If the object's semantic type is found to be in the candidate edge set, then we deem the predicate as valid.  In our experiments, we found that this filtering method significantly eliminates predicates which do not directly pertain to the biomedical domain. 
\else
\noindent \textbf{Semnet Filtering [Box 4]}. The raw predicates extracted from previous stage contain superfluous and spurious connections between concepts that can only serve to increase the noise in the training dataset. Hence we design two distinct filtering strategies to extract the most relevant predicates. Semnet filtering takes advantage of the UMLS concept's semantic types to construct a relationship graph between all known types. If the subject and object terms are connected via some path in this graph then the predicate is retained and otherwise removed.  
\fi 

\noindent \textbf{Co-Count Filtering [Box 5]}. A second strategy of pruning is to get information about the terms that co-occur the most in our corpus. A larger degree of co-occurrence implies a greater correlation between concepts. We use a normalized frequency count as a scoring measure of co-occurrence. All predicates that contain subject and object terms below a certain threshold are pruned from the list of candidate predicates. 

\section{Validation}
\label{sec:validation}

A fair validation of HG systems is extremely challenging, as these models are designed to predict \textit{novel} connections that are unknown to even those who evaluate the system~\cite{sybrandt2018b}. In addition, even if validated by rediscovering findings using historical, the process is computationally expensive because of the need to train multiple models to understand how many months (or years) back, the HG system can predict the findings which requires careful filtering of the used papers, vocabulary and other types of data. 
To present our results in terms of its usefulness for urgent CORD-19-related HG, we use a historical benchmark, which is conceptually described in~\cite{sybrandt2020agatha}. This technique is fully automated and does not require any domain experts intervention. 

\noindent{\bf Positive samples collection.}
We use \texttt{SemRep} and proposed pipeline in the previous section approach to process the most recent CORD-19 citations, which were published after the specific cut date making sure that the citations are not included in the training set. After that we extract all subject-object pairs from the obtained results and explicitly check that none of these pairs are presented in the training set. Pairs mentioned in the CORD-19 less than twice are filtered out from the validation set. Almost all of them are either noisy or represent information that already appears in other pairs (e.g., because of the difference in grammar).

We also use the strategy of {\bf subdomain recommendation}. This strategy works in the following way. For each UMLS term we collect its semantic type (which is a part of the metadata provided in UMLS metathesaurus) and group all extracted SemRep pairs by the term-pair criteria (combination of subject and object types). 
\ifthesis
    Then we identify the top-20 most common term-pairs subdomains and construct the validation set from pairs belonging to these 20 subdomains.
\else
    Then we identify the top-10 most common term-pairs subdomains and construct the validation set from pairs belonging to these 10 subdomains.
\fi

\noindent{\bf Negative samples generation.}
To generate negative samples per domain, the random sampling is used, that is, for each positive sample we keep its subject and randomly sample the object belonging to the same semantic type as the object of the source pair. We do this 10 times, thus having 10 negative domain-specific samples for each positive sample. 
When the validation set is generated, we apply our ranking criteria to it, obtaining a numerical score value $s$ per each sample, where $s \in [0, 1]$.

\noindent{\bf Evaluation metrics.} We propose our approach as a recommendation system and to report our results we use a combination of the following classification and recommendation metrics.
\begin{itemize}[noitemsep,topsep=0pt]
\item Classification metrics: (1) Area under the receiver-operating-characteristic curve (AUC ROC); (2) Area under the precision-recall curve (AUC PR).
\item Recommendation metrics: (1) Top-k precision (P.@k); (2) Average precision (AP.@k).
\end{itemize}
We report these numbers in per subdomain manner to better understand how the system performs with respect to specific task (e.g. drug repurposing). 
\section{Results}\label{sec:results}
To report results, we provide the performance measures for three \pymoliere{} models trained on the same input data (MEDLINE corpus and CORD-19 abstracts dataset):
\begin{enumerate}[noitemsep,topsep=0pt]
    \item \sysnameo : Baseline \pymoliere{} model~ \cite{sybrandt2020agatha};
    \item \sysnamec : \sysnameo with new UMLS layer and \texttt{SemRep} enrichment;
    \item \sysnamegp : \sysnamec with additional deep learning-based extracted and further filtered predicates.
\end{enumerate}
It is done in this particular manner because the major role in learning the proposed ranking criteria depends heavily on the quality of extracted semantic predicates and their number, as they form the training set for the \pymoliere{} ranking module. \emph{At the moment of writing, no other general purpose and available for public use HG system compliant with the three validation criteria, namely, (a) ability to run thousands of queries in a reasonable time, (b) ability to process COVID-19 related vocabulary, and (c) ability to operate in multiple domains was available for comparison.} Comparison of the baseline AGATHA-O is discussed in \cite{sybrandt2020agatha}.

The performance of both \sysnamec and \sysnamegp allows to run thousands of queries in a very short time (in the order of minutes), making the validation on a large number of samples possible. Unfortunately, given the current circumstances, large-scale validation for the specific scientific subdomain (COVID-19 related hypotheses) is hard to implement, because well-established and reliable factual base is being actively developed at the moment and big historic gap for the vocabulary simply does not exist (e.g., the COVID-19 term is less than two years old). We, however, provide the validation set of positive connections extracted from CORD-19 dataset citations added within the time frame from October 28, 2020 to January 21, 2021, which numbered at 77 thousand abstracts.

\ifthesis
    The approximate running time with corresponding types of used hardware  is presented in Table \ref{tab:running_times2}. Each row corresponds to the stage in the \sysnamec/\sysnamegp pipelines. The column ``M'' (machines) and CPU show the number of machines and required CPUs, respectively. In the column ``GPU'' we indicate if GPU was required or optional. For AGATHA training we used two NVIDIA V100 per machine. The minimal requirements for RAM per machine are in column ``RAM''. The running time of queries is negligible.


 
 

\begin{table}
\caption{
      \label{tab:running_times2}
      Running time and hardware requirements.
    }
\begin{tabular}{llllll}
\toprule
 Stage                                   &        Time       & \multicolumn{4}{c}{Hardware}                           \\
\midrule
                               & & M & CPU & GPU          & RAM \\
\midrule
SemRep Processing            & 2 d       & 10-28       & 20+            & Opt  & N/A                \\
AllenNLP Predicates      &   3 d &  28-40        & 20+             &    Opt            &       N/A       \\
Graph Construction         & 10 d    & 30+         & 20+            & Opt  & 120GB+            \\
Graph Conversion             & 7 h    & 1           & 40+            & Opt  & 1TB+              \\
Graph Embedding                & 1 d        & 20          & 24+            & Opt  & 120GB+            \\
AGATHA Training       & 22 h  & 5+          & 2+             & Yes & 300GB+            \\
Network Adjacency & 1 d        & 1           & 40+            & Opt  & 1.5TB+ \\
\bottomrule
\vspace*{2mm}
\end{tabular}
\end{table}
    
    \iffalse
        \input{tables/graph_metrics}
        In Table~\ref{tab:graph_metrics}, we share some basic graph metrics for the models \sysnameo, \sysnamec and \sysnamegp. The most significant change is observed in the number of semantic predicates and coded terms, which clearly represents the purpose of introducing additional preprocessing steps. 
        \else 
        The overall training dataset contains 190.6 million sentences, which results in 287 million nodes and 13.5 billion edges (\sysnamec model).
    \fi
\fi

\begin{table*}[]

    \centering
    \caption{
    Classification and recommendation quality metrics across recently popular COVID-19-related biomedical subdomains. Labels O, C and GP stand for \sysnameo, \sysnamec and \sysnamegp models, respectively. Used abbreviations: 
        \textit{phsu}: Pharmacologic Substance;
        \textit{topp}: Therapeutic or Preventive Procedure;
        \textit{humn}: Human;
        \textit{menp}: Mental Process;
        \textit{spco}: Spatial Concept;
        \textit{aapp}: Amino Acid, Peptide, or Protein;
        \textit{gngm}: Gene or Genome;
        \textit{orch}: Organic Chemical;
        \textit{fndg}: Finding;
        \textit{geoa}: Geographic Area;
        \textit{dsyn}: Disease or Syndrome.
    Semantic pairs are presented in descending order according to the citations count of their corresponding predicates.
    }
    \setlength{\tabcolsep}{.35em}
    \renewcommand{\arraystretch}{1.1}
    \begin{tabular}{lccccccccccccccccccccc}
\toprule
{} & 
\multicolumn{3}{c}{ROC AUC} &
\multicolumn{3}{c}{PR AUC} & 
\multicolumn{3}{c}{P.@10} & 
\multicolumn{3}{c}{P.@100} & 
\multicolumn{3}{c}{AP.@10} & 
\multicolumn{3}{c}{AP.@100} \\

\cmidrule(lr){2-4}
\cmidrule(lr){5-7}
\cmidrule(lr){8-10}
\cmidrule(lr){11-13}
\cmidrule(lr){14-16}
\cmidrule(lr){17-19}
{} &       
O &     C &     GP &   
O &     C &     GP &   
O &     C &     GP &      
O &     C &     GP &       
O &     C &     GP &     
O &     C &     GP \\[-3pt]
\midrule
dsyn:dsyn &    0.83 &  0.89 &  0.88 &   0.35 &  0.41 &  0.44 &   0.60 &  0.50 &  0.70 &    0.54 &  0.54 &  0.55 &    0.70 &  0.49 &  0.72 &      0.56 &  0.54 &  0.62 \\[-3pt]
phsu:dsyn &    0.86 &  0.91 &  0.91 &   0.36 &  0.43 &  0.47 &   0.30 &  0.60 &  0.60 &    0.45 &  0.51 &  0.57 &    0.77 &  0.73 &  0.90 &      0.47 &  0.55 &  0.65 \\[-3pt]
fndg:dsyn &    0.86 &  0.93 &  0.92 &   0.42 &  0.53 &  0.57 &   0.70 &  0.70 &  1.00 &    0.54 &  0.58 &  0.66 &    0.90 &  0.70 &  1.00 &      0.65 &  0.64 &  0.78 \\[-3pt]
dsyn:fndg &    0.80 &  0.89 &  0.90 &   0.29 &  0.42 &  0.43 &   0.40 &  0.60 &  0.60 &    0.39 &  0.46 &  0.49 &    0.46 &  0.86 &  0.55 &      0.44 &  0.59 &  0.54 \\[-3pt]
fndg:humn &    0.80 &  0.89 &  0.89 &   0.37 &  0.46 &  0.47 &   0.90 &  0.80 &  0.80 &    0.43 &  0.54 &  0.58 &    0.96 &  0.86 &  0.70 &      0.71 &  0.66 &  0.68 \\[-3pt]
dsyn:humn &    0.77 &  0.84 &  0.85 &   0.28 &  0.32 &  0.35 &   0.50 &  0.30 &  0.40 &    0.35 &  0.41 &  0.47 &    0.59 &  0.33 &  0.30 &      0.53 &  0.44 &  0.48 \\[-3pt]
topp:dsyn &    0.87 &  0.92 &  0.92 &   0.38 &  0.50 &  0.50 &   0.40 &  0.80 &  0.60 &    0.51 &  0.58 &  0.61 &    0.33 &  0.70 &  0.81 &      0.48 &  0.65 &  0.63 \\[-3pt]
orch:dsyn &    0.87 &  0.90 &  0.88 &   0.35 &  0.45 &  0.46 &   0.50 &  0.80 &  0.60 &    0.38 &  0.43 &  0.53 &    0.69 &  0.96 &  0.91 &      0.47 &  0.63 &  0.63 \\[-3pt]
geoa:spco &    0.74 &  0.72 &  0.90 &   0.22 &  0.19 &  0.43 &   0.30 &  0.20 &  0.60 &    0.31 &  0.23 &  0.53 &    0.41 &  0.37 &  0.68 &      0.32 &  0.24 &  0.53 \\[-3pt]
aapp:dsyn &    0.87 &  0.93 &  0.92 &   0.39 &  0.47 &  0.50 &   0.40 &  0.40 &  0.70 &    0.47 &  0.51 &  0.53 &    0.52 &  0.28 &  0.88 &      0.50 &  0.50 &  0.60 \\[-3pt]
aapp:gngm &    0.73 &  0.85 &  0.84 &   0.20 &  0.30 &  0.34 &   0.30 &  0.00 &  0.60 &    0.20 &  0.36 &  0.36 &    0.32 &  0.00 &  0.59 &      0.32 &  0.33 &  0.44 \\[-3pt]
orch:orch &    0.86 &  0.92 &  0.91 &   0.54 &  0.65 &  0.61 &   0.80 &  0.90 &  0.80 &    0.53 &  0.58 &  0.56 &    0.99 &  0.99 &  0.88 &      0.73 &  0.81 &  0.75 \\[-3pt]
gngm:dsyn &    0.91 &  0.95 &  0.94 &   0.44 &  0.65 &  0.53 &   0.70 &  0.80 &  0.50 &    0.43 &  0.56 &  0.57 &    0.87 &  0.95 &  0.69 &      0.52 &  0.75 &  0.59 \\[-3pt]
orch:phsu &    0.87 &  0.92 &  0.93 &   0.52 &  0.61 &  0.59 &   0.80 &  0.80 &  0.60 &    0.49 &  0.55 &  0.58 &    0.94 &  0.95 &  0.90 &      0.70 &  0.73 &  0.67 \\[-3pt]
gngm:gngm &    0.71 &  0.85 &  0.85 &   0.18 &  0.35 &  0.34 &   0.30 &  0.60 &  0.40 &    0.20 &  0.33 &  0.36 &    0.35 &  0.65 &  0.35 &      0.22 &  0.47 &  0.45 \\[-3pt]
phsu:fndg &    0.76 &  0.88 &  0.87 &   0.24 &  0.45 &  0.44 &   0.40 &  0.70 &  0.80 &    0.29 &  0.44 &  0.42 &    0.30 &  0.69 &  0.88 &      0.34 &  0.58 &  0.57 \\[-3pt]
phsu:phsu &    0.86 &  0.91 &  0.92 &   0.53 &  0.56 &  0.58 &   0.90 &  0.80 &  0.80 &    0.45 &  0.47 &  0.53 &    0.98 &  0.82 &  0.90 &      0.72 &  0.70 &  0.69 \\[-3pt]
orch:aapp &    0.90 &  0.93 &  0.93 &   0.57 &  0.61 &  0.68 &   0.80 &  0.70 &  1.00 &    0.49 &  0.53 &  0.58 &    0.96 &  0.74 &  1.00 &      0.74 &  0.71 &  0.78 \\[-3pt]
geoa:menp &    0.88 &  0.89 &  0.94 &   0.40 &  0.50 &  0.61 &   0.60 &  0.60 &  0.60 &    0.42 &  0.47 &  0.58 &    0.59 &  0.84 &  0.59 &      0.47 &  0.63 &  0.68 \\[-3pt]
phsu:orch &    0.88 &  0.94 &  0.94 &   0.49 &  0.62 &  0.55 &   0.70 &  0.80 &  0.60 &    0.44 &  0.48 &  0.51 &    0.92 &  0.95 &  0.92 &      0.62 &  0.74 &  0.61 \\[-3pt]
\bottomrule
Mean      &    
0.83 &  0.89 &  \textbf{0.90} &   
0.38 &  0.47 &  \textbf{0.49} &   
0.57 &  0.62 &  \textbf{0.66} &    
0.42 &  0.48 &  \textbf{0.53} &    
0.68 &  0.69 &  \textbf{0.76} &      
0.53 &  0.59 &  \textbf{0.62} \\[-3pt]

\end{tabular}
    
    \label{tab:models_big_comparison_aaai}
\end{table*}

In Table~\ref{tab:models_big_comparison_aaai}, we compare aforementioned models using the metrics described in the previous section. We present predicate types with NLM semantic type codes~\cite{pmid11604736} due to space restrictions. \textit{Both \sysnamec and \sysnamegp models  show significant gains when compared to \sysnameo baseline model}. Benefits in the most problematic for the baseline model areas (e.g., \textit{(Geographic Area) $\rightarrow$ (Spatial Concept)} denoted by \textit{(geoa,spco)}) serve the best illustration for that, showing up to 0.25 advantage in ROC AUC (\sysnamegp). Important biomedical domains, such as \textit{(Amino Acid, Peptide or Protein) $\rightarrow$ (Disease or Syndrome)} denoted by \textit{(aapp,dsyn)} also show noticeable improvements (0.07 for \sysnamec\nolinebreak). Due to space limit we include the results for only top-10 most popular subdomains. More can be  found in the long preprint version~\cite{tyagin2021accelerating}.
Average ROC AUC value is increased by 0.07.

Our validation strategy involves a big number of many-to-many queries, making the area under precision-recall curve another very illustrative metric. This is where the newly proposed models show even more drastic improvements over the baseline \sysnameo. For some subdomains, like \textit{(Organic Chemical) $\rightarrow$ (Disease or Syndrome) (orch,dsyn)} we observe that new models improve the PR AUC score on more than 0.1. Average PR AUC value is increased by 0.12.

\section{Emergent discovery case study}
\label{sec:case_study}

The proactive discovery of ongoing research findings is an important component in the validation of hypothesis generation systems \cite{sybrandt2018a}. In particular, in the current uncertain situation with COVID-19 when a lot of unintentionally incorrect discoveries are published, the validation must include human-in-the-loop part even in limited capacity such as in \cite{aksenova2019inhibition,spangler2014automated}. To demonstrate the predictive potential of \sysnamec we perform a case study on three COVID-19-related novel connections manually selected by the domain expert. These connections were published after the cut date before which any data used in training was available to download at NIH. 

\begin{figure*}
    \centering
    \includegraphics[
        width=1\linewidth,
    ]{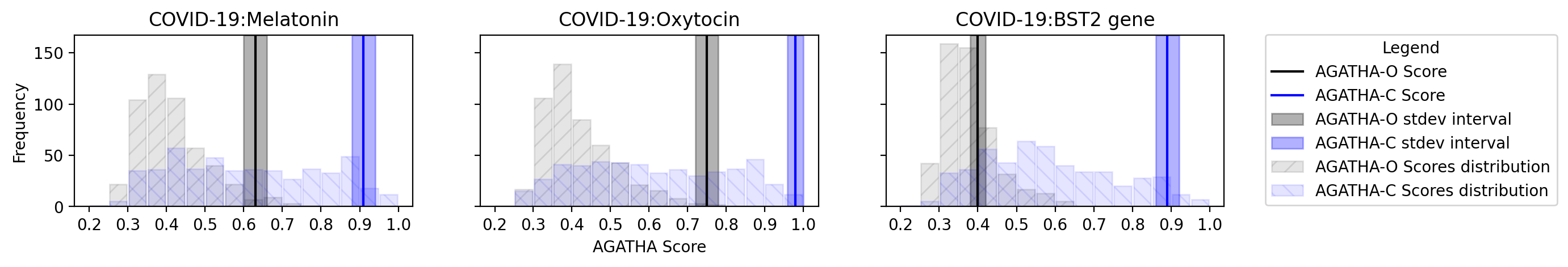}
    \caption{
      \label{fig:case_study_histograms}
      Score distributions in case study experiment. Presented scores are obtained with \sysnameo and \sysnamec models.  
    }
\end{figure*}

At a low level, all \pymoliere{} models use entity subsampling to calculate pairwise ranking criteria, which means that the absolute numbers may fluctuate slightly. Thus, to present the numeric scores, each experiment was repeated 100 times to compute the average and standard deviation that we present in Figure \ref{fig:case_study_histograms}.

\sysnamec was tested whether it will be able to predict compounds potentially applicable for the treatment of COVID-19 and the genes involved in the SARS-CoV-2 pathogenesis. The data confirming  cardiovascular protective effects of hormone oxytocine were published recently 
\cite{diep2021there,wang2021cardiovascular}. The protective effect is linked to anti inflammatory activity of the hormone. \sysnamec ranked this connection at top 1.4 percent.

Similarly, we tested the prediction of the effects of the other hormone, melatonin. Several publications, started from November 2020 \cite{cardinali2020can,zimmermann2020covid,alschuler2020integrative,ho2021perspective} show the protective effects of melatonin, specifically for COVID-19 neurological complications. The activity was linked to anti-oxidative effects of the melatonin. This connection was ranked at top 5.6 percent.

Our system ranked at top 7.6 percent the involvement of  tetherin (BST2). The results published in 2021 \cite{stewart2021sars} show that tetherin restricts the secretion of SARS-CoV-2 viral particles and is downregulated by SARS-CoV-2. Therefore, pharmacological activation of tetherin expression, or inhibition of the degradation could be a promising direction of the development of SARS-CoV-2 treatment.

To demonstrate AGATHA-C ranking capabilities, we use similar strategy to what we proposed in
the validation section, but now we randomly generate 500 negative samples for each pair of interest, maintaining the ratio of 1:500 between real-world connections and random noise. The goal of this experiment is to rank the \emph{real} connections  above randomly sampled pairs of the same semantic network types. Illustration of this experiment is presented in Figure \ref{fig:case_study_histograms}. In each of the  three sub-figures, there are 501 results of scores for 501 pairs. One of the results is now known to be correct, while other 500 are unknown. For example, for the first known meaningful pair COVID-19-melatonin, we generated 500 pairs COVID-19-X pairs, where X is a randomly chosen hormone. Because we are dealing with a medical domain, nobody can be 100\% sure that all these 500 are irrelevant. However, most of them are not likely to be related to COVID-19. In the histograms of Fig. \ref{fig:case_study_histograms} we show a distribution of scores for each 501 experiments for two models: baseline \sysnameo and newly proposed \sysnamec\nolinebreak. The solid lines indicate actual scores for the known pairs. To conclude, if lab experiments were required to confirm the connections, the known connections would be confirmed in the top 10 percent of predictions which significantly reduces the research time and cost.

\begin{table}[h]

    \centering
    
    \caption{
        \label{tab:aug_model_comp}Comparison between \sysnamec models trained with different cut dates and validated with recently discovered pairs. 
    }

    \begin{tabular}{rcccc}
        
        \toprule
        {Model} &  ROC AUC &  PR AUC &  AP.@10 \\
        \midrule
        C (Oct 2020) &     0.88 &    0.49 &    0.67 \\
        C (June 2021)    &     0.90 &    0.55 &    0.77 \\
        \bottomrule
   
    \end{tabular}

\end{table}

We also test how inclusion of more recent training data affects AGATHA performance. For that we take two models trained with the same methodology (C models), but one model contains only training data limited to October 28th 2020 and another model is larger and contains more recent data (threshold: June 23rd 2021). Both models were validated with the pairs firstly introduced between June 24th 2021 and August 11th 2021. Results of this experiment presented in Table \ref{tab:aug_model_comp}. It shows that retraining the model using the proposed method and more recent data yields in slightly better scores in all basic metrics.


\section{Lessons Learned and Open Problems}
\label{sec:lessons_learned}

\noindent{\bf Quality of the information retrieval pipelines}. Information retrieval is an important part of any HG pipeline. In order to uncover \textit{implicit} connections, the system should be able to capture existing \textit{explicit} connections with as much quality as possible. 
\ifthesis
    Given that human knowledge is usually stored in a non-structured manner (e.g.,  scientific texts), the quality of systems that process raw textual data, such as those that solve the named entity recognition, or word sense disambiguation problems,
    is crucial.
\fi

We observed that the SemRep system performs better concept and relation recognition when full abstracts are used as input data instead of single sentences. SemRep also allows to perform optional sortal anaphora resolution to extract co-references to the entities from neighbouring sentences, which was shown to be useful in~\cite{semrep2016anaphora} and is used in this work.

\noindent{\bf ``Positive" research bias}. The absence of published negative research results is a big problem for the HG field. With mostly positive results available, often we have to generate negative examples through some kind of random sampling. These negative samples likely do not adequately represent the real nature of negatively confirmed scientific findings. Likely, one of the most important future work directions in the area of HG is to accurately distinguish and leverage positive and negative proposed results.

\ifthesis
    \noindent{\bf Domain experts involvement}. When any hypothesis generation system is built, one of the first questions a designer should address is extent that domain experts are expected to participate in the pipeline. Modern decision-making systems allow a fully automated discovery process (like the \pymoliere{} system), 
    but this may not be sufficient.
    A domain expert
    who interfaces with a HG system as a black box may not trust generated results or know how best to interpret them.
    The challenge of interpretable hypothesis generation remains a significant barrier to widespread adoption of these kinds of research tools.
    For this we advocate using our ``structural'' learning HG system MOLIERE \cite{sybrandt2017} in which with the topical modeling and network analytic measures we interpret and explain the results.
\fi

\noindent{\bf The nature of input corpora}. The question of what should be used as input to a topic-modeling based hypothesis generation system is raised in \cite{sybrandt2018b}. Using full-text papers shows an improvement, but the trade-off between run time and output quality was barely justifiable. However, deep learning models have a greater potential for extracting useful information from large input sources, and as it was demonstrated in our previous work~\cite{sybrandt2020agatha}, show significant performance advancements. Thus the question of using full-text papers in deep learning-based hypothesis generation systems should be addressed. 

\noindent{\bf Knowledge resolution}. Our newly proposed systems showed that the knowledge resolution plays a major role in subdomain recommendation. To increase the scope of model expertise (and the scope of potential applications beyond the biomedical fields) we deliberately incorporate
a general-purpose information retrieval system RnnOIE into \sysnamegp.
\ifthesis
    This additional information results in significant gains in broad subdomains like \textit{(Geographic Area) $\rightarrow$ (Idea or Concept) (geoa,idcn)}.
    At the same time, we observe that \sysnamec performs better in ``microscopic'' biomedical areas, e.g. \textit{(Organic Chemical) $\rightarrow$ (Organic Chemical) (orch,orch)}, which raises the question of choosing the appropriate model for every specific use case. 
\fi
Although, both systems process all types of queries, the general purpose predicates participated in training significantly improve ``macroscopic'' types of queries.

\noindent{\bf Predicate Extraction}. One of the most important aspects of any hypothesis generation system is to give it the ability to \emph{reject} hypothesis which are not backed by any research. This task becomes difficult when we consider the positive research bias of the existing literature. 
\ifthesis 
    One way to approach this problem can be to integrate relation extraction with sentiment modeling. In this way, we would be able to capture not only the relations but also \emph{how} they are connected. Some work~\cite{zhou_span-based_2019, eberts_span-based_2019} has been proposed in this direction, but it remains an active area of research in the biomedical domain. We aim to address this enhancement in a future work. In our experiments, we found the LSTM based RnnOIE model to be one of the most computationally demanding components of our pipeline. Speeding up inference time and improving data processing efficiency at scale are two open problems in this area.
\else
    We aim to address this enhancement in a future work.

\fi

\ifthesis
    \noindent{\bf Deployment Strategy}. It is not easy to define a perfectly correct result in biomedical hypothesis generation domain. Every year well known drugs (sometimes used for generations) are retracted from the market because of the newly discovered adversarial reactions, side effects or other contradictions. Although, our system is publicly available, we cannot claim that it is fully deployed because a full deployment with massive indication of good results should always include a large series or proactive discovery  supported with the lab experiments. Thus, a dissemination of such system in academic biomedical labs, and pharmaceutical research companies is imperative. Specifically, important component of the full deployment that we are unable to test without such broad collaborations is a verification of proprietary experimental failure results as it is known that for a single working drug main molecule, there are many more of  those that have not been experimentally confirmed to be useful. This negative experimental information is usually not published but documented in internal proprietary journals.

    \noindent{\bf Emergent vs. traditional research.} Emergent research may show results, which differ significantly from the results obtained with well-established and widely used methods and datasets. One particularly important aspect of emergent research (such as the one that we observe during COVID-19 pandemic) is the validation strategy. When we started to work on the system there was no publicly available high quality sufficiently large validation set, which could have been used to evaluate the system at scale because COVID-19 related terms have not been mentioned enough with other different terms. For the evaluation purposes we use "time slicing" technique in which we  keep the most novel research out of our training set. The problem with this approach is that in some cases there is not enough time for a novel research to get proven or refuted by a scientific community. This is especially sensitive issue in the biomedical research.  For example, in the original AGATHA model  demonstration~\cite{sybrandt2020agatha} we used a more sophisticated strategy to construct the validation set, which included finding the most popular connections with non-decreasing number of citations over time. It was possible because the validation set was large enough to accommodate strict filtering and extracting only a fraction of results. In this paper we had to significantly loosen these criteria to keep the dataset size as large as possible to represent the overall system performance.
    
\fi
\section{Conclusions}
\label{sec:conclusion}

We present two graph mining transformer based models  \sysnamec and \sysnamegp, for micro- and macroscopic scales of queries respectively, which are designed to help domain experts solve high-priority research problems and accelerate scientific discovery. We perform per-subdomain validation of these new models on a rapidly changing COVID-19 focused dataset, composed of recently published concept pairs and demonstrate that the proposed models achieve state-of-the-art prediction quality. Both models  significantly outperform the existing baselines. We deploy the proposed models to the broad scientific community and believe that our contribution can raise more interest in prospective hypothesis generation applications.


\bibliographystyle{ACM-Reference-Format}
\bibliography{main}


\ifthesis
\else
\pagebreak

\ifthesis
\section*{Reproducability Details}

\noindent{\bf Training Graph Embedding}
When optimizing our semantic graph embedding, we find that maximal performance
is achieved using a compute cluster of twenty twenty-four-core machines. Within
the 72h time restriction of the Palmetto super computing cluster, we have
enough time to see every edge in the graph 10 times, in the case of the
256-dim embedding, and 5 times in the case of the 512-dim
embedding. Once complete, we are ready to begin training the \pymoliere{}
deep learning hypothesis generation model.

\noindent{\bf Training Ranking Model}
We minimize the ranking loss over
all published predicates using the LAMB optimizer \cite{you2019large}. This
allows us to efficiently train using very large batch sizes, which is necessary
as we leverage 10 NVIDIA V100 GPUs to effectively process 600 positive samples
(and therefore 2,400 total samples) per batch. In terms of hyperparameters, we
select a learning rate of $\eta=0.01$ with a linear warm up of 1,000 batches, a
margin of $m=0.1$, a neighborhood sub-sampling rate of $s=15$, and we perform
cross-validation on a $1\%$ random holdout to provide early stopping and to
select the best model with respect to validation loss. Due to the large size of
training data, one epoch consists of only $10\%$ of the overall training data.
This process is made easier through the helpful Pytorch-Lightning
library~\cite{falcon2019lightning}.

\input{tables/model_parameter_count.tex}
\input{tables/graph_size.tex}

\fi

\fi

\end{document}